\documentclass[desactivate]{aa}
\usepackage{graphicx}
\usepackage{txfonts}
\usepackage{placeins}
\usepackage{float}

\usepackage{nameref}

\begin{document}

\titlerunning{Exploring the effects of diffuse ionised gas in UM 462 and IIZw40}
\authorrunning{Lagos et al.}

   \title{Exploring the effects of diffuse ionised gas in two local analogues of high-redshift star-forming galaxies}



   \author{P. Lagos\inst{1,2}
              \and A. Nigoche-Netro\inst{3}
          \and T. C. Scott\inst{2}
          \and C. Sengupta\inst{4,5}
          \and R. Demarco\inst{1}
          }
\institute{Institute of Astrophysics, Facultad de Ciencias Exactas, Universidad
	       Andr\'es Bello, Sede Concepci\'on, Talcahuano, Chile
        \and
            Instituto de Astrofísica e Ciências do Espaço, Universidade do Porto, CAUP, Rua das Estrelas, 4150-762 Porto, Portugal\\
              \email{Patricio.Lagos@astro.up.pt}
         \and         
         Instituto de Astronomía y Meteorología, Universidad de Guadalajara, Guadalajara, Jal. 44130, México
         \and
         Centre for Space Research, North-West University, Potchefstroom 2520, South Africa
         \and National Institute for Theoretical and Computational Sciences (NITheCS), Potchefstroom 2520, South Africa\\         
         }

   \date{Received September 15, 1996; accepted March 16, 1997}

 
  \abstract
   {}
   {
    We investigate the impact of diffuse ionised gas (DIG) on the determination of
    emission line ratios and gas-phase metallicities in two local analogues of
    high-redshift star-forming galaxies: UM 462 and IIZw 40. Understanding how DIG
    affects these quantities is essential for interpreting unresolved observations of
    distant galaxies, where integrated spectra are often used to trace their chemical
    evolution.}
   {
    Using archival Very Large Telescope, Multi-Unit Spectroscopic Explorer
    (MUSE) data, we spatially resolved the warm ionised medium of both galaxies. 
    We derived oxygen abundances through the direct method and several HII-based
    strong-line calibrators, and we used the H$\alpha$ surface brightness
    ($\Sigma$(H$\alpha$)) to distinguish regions dominated by HII or DIG emission.}
   {
    Oxygen abundances derived from the N2 and O3N2 indices show an inverse
    correlation with $\Sigma$(H$\alpha$), ionisation parameters,
    and H$\alpha$ equivalent width (EW(H$\alpha$)), with DIG-dominated regions
    exhibiting 12 + log(O/H) values higher than the mean for their galaxy by
    $\sim$0.2 dex in UM 462 and $\sim$0.1 dex in IIZw40. The metallicity differences
    derived from these strong-line calibrators reach about 0.4 dex and 0.3 dex
    between the highest (HII-dominated) and lowest (DIG-dominated) 
    $\Sigma$(H$\alpha$) bins in UM 462 and IIZw40, respectively.
    We found a linear correlation between $\Delta(\text{O/H})$ 
    (O/H deviation from the mean interstellar medium value) and EW(H$\alpha$). 
    Trends with $\Sigma$(H$\alpha$), metallicity, EW(H$\alpha$), 
    and ionisation parameter suggest smoothly evolving ionisation conditions  
    in the interstellar medium in our galaxies.
    Such trends and metallicty variations derived from HII-based calibrators reflect
    different ionisation sources and levels rather than true abundance changes. 
    In particular, the use of these calibrators can lead to spurious
    metallicity gradients in galaxies with extended DIG tails, such as tadpole or
    cometary-like galaxies, which can be misinterpreted as evidence of the infall of
    metal-poor gas. The most likely mechanism for ionising the DIG in our sample of
    HII or BCD galaxies is the leakage of photons from HII regions, with shocks
    induced by stellar feedback processes also contributing significantly. 
    Consequently, such contamination may affect the reliability of the derived oxygen
    abundances. Our results highlight the importance of accounting for DIG 
    in galaxy metallicity estimates since it potentially biases metallicity gradient
    measurements. This is particularly relevant for large surveys at high redshift
    that rely on integrated galaxy spectra.
  }  
   {}
  

   \maketitle
%

\section{Introduction}

Local star-forming galaxies, such as dwarf HII galaxies and/or blue compact
dwarfs (BCDs), are critical for understanding the processes that shaped galaxies in
the early Universe. These nearby systems, with their intense star formation,  
low mass, and low metallicities, closely resemble galaxies that existed in the early
Universe. By studying these local systems, we can better interpret the
physical conditions and evolutionary processes of their high-redshift counterparts.
Notably, such investigations create a vital connection between the detailed phenomena
observable in the local Universe and the broader picture over cosmic time of galaxy
formation revealed by large ground- and space-based telescopes, such as the
\textit{Hubble} Space Telescope \citep[HST; e.g.][]{Beckwith2006} and more recently
the \textit{James Webb} Space Telescope \citep[JWST; e.g.][]{Sanders2023}.

The warm ionised component of the interstellar medium (ISM) in star-forming galaxies
can be divided into two main components: HII regions, where hydrogen is fully ionised
by young and massive stars, and the diffuse ionised gas (DIG), a more extended,
lower-density component outside of HII regions characterised by lower
H$\alpha$ surface brightness ($\Sigma$(H$\alpha$)). 
There are several likely mechanisms responsible for ionising the DIG, including
(i) Lyman escape photons from the HII regions \citep[e.g.][]{,Giammanco2004},
(ii) shock ionisation driven by  stellar winds and/or outflows
\citep[e.g.][]{DopitaSutherland1995,Tullman2000,Allen2008,LopezCoba2020} 
in star-forming galaxies, (iii) photoionisation from hot low-mass evolved 
stars (HOLMES) or post-asymptotic giant branch (post-AGB) stars in retired galaxies
and regions within them
\cite[e.g.][]{Stasinska2008,Singh2013,Lacerda2018,Lagos2022}, (iv) turbulent mixing
layers \cite[][]{Rand1998ApJ,Binette2009}, and (v) cosmic rays \cite[][]{Wiener2013}, 
among others. Several methods have been developed and used to segregate HII- and
DIG-dominated regions or to account for DIG contamination
\citep[e.g.][]{Kaplan2016,Lacerda2018,Sanchez2020,Sanchez2021, Metha2022,Lugo-Aranda2024}. 
However, the broader impact of these two regimes on galaxy properties at different 
redshifts remains unclear \citep[e.g.][]{Zhang2017,Tomicic2021}. 

The DIG affects the emission line ratios such as
[NII]/H$\alpha$, [SII]/H$\alpha$, and [OIII]/H$\beta$, which tend to increase
with distance from the galaxy centre, thereby influencing the inferred gas-phase
metallicity and the corresponding gradients obtained from strong-line
methods \citep[e.g.][]{ValeAsari2019}.
Strong-line methods \cite[e.g.][]{Pettini2004,Marino2013}, which are used to derive
metallicity, are calibrated using the direct method (measurement of the electron
temperature T$_e$), photoionisation models \cite[e.g.][]{Kewley2002}, or a combination
of both \cite[e.g.][]{Denicolo2002}.  
Moreover, the stellar mass--star formation rate relation \cite[e.g.][]{Mannucci2010}
should be interpreted with care, as the star-formation rate derived from L(H$\alpha$)
may be biased due to DIG contamination.
These emission line ratios are also affected by the relative strength of the DIG and
HII emission \citep{Poetrodjojo2019}, but observations usually lack sufficient
resolution to clearly distinguish DIG- from HII-dominated regions.
 
The ionised gas in HII galaxies and BCDs show complex and, in some cases, turbulent
kinematic features \citep[e.g.][]{Bordalo2009,Moiseev2012} as well as ﬁlamentary
structures, shells, and cavities resulting from the kinetic energy of stellar winds 
and supernova explosions and features arising from flyby interactions
\cite[e.g.][]{Scott2024}, which makes distinguishing and separating the HII- and
DIG-dominated regions very challenging. In fact, most of the stellar tails in our
sample of cometary and/or extremely metal-poor (XMP; 12 + log(O/H) $\lesssim$ 7.6)
BCDs  \citep{Papaderos2008} are co-spatial with an extended DIG component
\citep{Lagos2016,PapaderosOstlin2012,Lagos2014,Lagos2018}. 
The presence of DIG may significantly influence the physical and chemical properties
of these galaxies.

In this paper we use archival Very Large Telescope (VLT), Multi-Unit Spectroscopic
Explorer (MUSE) data from two local HII and/or BCD galaxies (UM 462 and IIZw40) 
to investigate these issues and gain insights into the  processes ionising the DIG
and their impact on integrated galaxy properties. 
The MUSE spatial sampling is $\sim$12 pc and $\sim$10 pc
per pixel or spaxel for UM 462 and IIZw40, respectively, which is sufficient to
distinguish between DIG- and HII-dominated regions.
The structure of this article is as follows: Section \ref{Sect_observations}
presents the observations, data reduction, and spectral fitting used in this study. 
In Sect. \ref{Sect_results} we describe our results as well as the DIG definition, the
determination of emission line ratios, ionisation parameters, and oxygen abundances. 
Finally, in Sect. \ref{Sect_discussion} we present a discussion and summarise our
conclusions.


\section{Observations, data reduction, and spectral fitting}\label{Sect_observations}%

Our two BCD galaxies are from \cite{Lagos2007}, and they were selected for their
extended ionised gas regions and differences in the strength of the current
starburst. IIZw40 is undergoing a strong starburst arising from an ongoing merger, 
while UM\,462 is relatively unperturbed \citep{vanZee1998A}. 
Both galaxies have archival ESO VLT, MUSE \citep{Bacon2010} data. UM 462 was
comprehensively analysed by \cite{MonrealIbero2023}, while \cite{Marasco2023}
analysed only  the kinematics of UM 462 and IIZw40. The corresponding ID programs are
0101.A-0282(A) for UM 462 and 094.B-0745(A) for IIZw40. 
The MUSE observations  were made in wide-field mode, with a 1' × 1' field of view.
This setup covers a spectral range of $\sim$4800--9300 $\AA$, with 
spatial and spectral sampling of 0.2" per spaxel and 1.25 $\AA$, respectively. 
Offset sky observations enabled us to perform effective sky subtraction.
Table \ref{Table_1} summarises the galaxy parameters and observing log.
Data were processed using the MUSE pipeline \cite[v2.6.2;][]{Weilbacher2020}, which
included bias correction, wavelength calibration, cube assembly, heliocentric
correction, sky subtraction, and exposure merging. The resulting data cubes were
further corrected  for redshift and Galactic extinction using the reddening law from
\cite{Cardelli1989}. See \cite{Lagos2022} for further details.

\begin{table}[H]
        \caption{General parameters of our sample galaxies and observing log.}
        \centering
        \begin{tabular}{lcc}
                \hline\hline
                & UM 462         & IIZw40\\
                \hline
                RA(J2000)  & 11h52m37.1926s & 05h55m42.6147s\\
                Dec(J2000) &-02d28m09.907s  &+03d23m31.657s\\
                A$_V$\tablefootmark{a} (mag)& 0.127 & 2.260 \\
                D (Mpc)    & 12.55          & 10.33\\
                Scale (kpc/arcsec) & 0.061 & 0.050 \\
                z          & 0.003469       & 0.002682\\
                12+log(O/H)& 8.02\tablefootmark{b} & 8.07\tablefootmark{c}\\
                \hline
                Date of observation & 2018-04-17 & 2014-12-29 \\
                Exp. time & 4x500 s & 3x960 s \\
                Mean airmass & 1.08 & 1.22\\
                \hline
        \end{tabular}
        \tablefoot{The top rows show the general parameters of our sample from
                NED. The lower rows contain observing log information.\\
                \tablefoottext{a}{Galactic extinction from \cite{Schlafly2011}.
                \tablefoottext{b}{\cite[][]{MonrealIbero2023}}.
                \tablefoottext{c}{\cite[][]{Bordalo2011}}.
                }
        }
        \label{Table_1}%
\end{table}

The stellar continuum  was removed based on the spectral synthesis code \textsc{FADO}
\citep{Gomes2017} version V1.B, with SSP templates from \cite{BruzualCharlot2003},
metallicities Z = 0.004-0.05, and stellar ages from 1 Myr to 15 Gyr for a Chabrier 
initial mass function \citep[IMF,][]{Chabrier2003}. 
The best-fit stellar continuum was subtracted, and each emission line was subsequently
fitted with a single Gaussian profile to generate the emission line maps.
More details are provided in \cite{Lagos2022}. 
Pixels and/or spaxels in the maps with a signal-to-noise ratio (S/N) of less than
five were excluded from the analysis.

\section{Results}\label{Sect_results}%

\subsection{H$\alpha$ emission line, EW(H$\alpha$), and $\sigma$(H$\alpha$)}

\begin{figure*}[h]
   \centering
   \includegraphics[width=6cm]{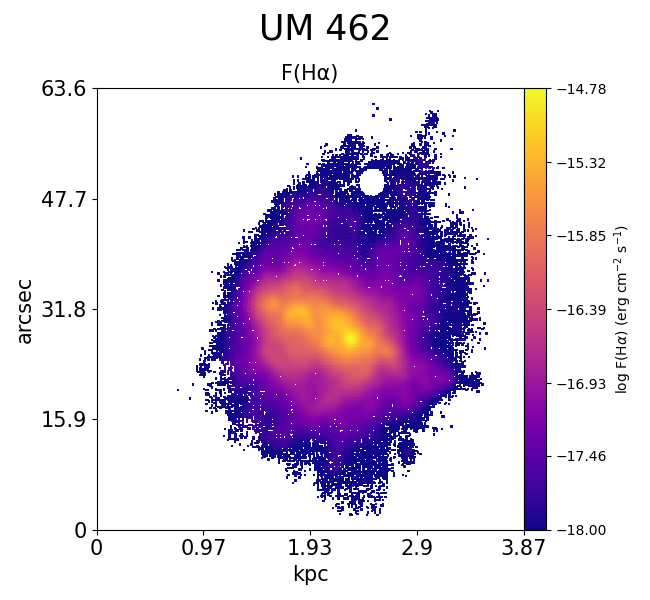}
   \includegraphics[width=6cm]{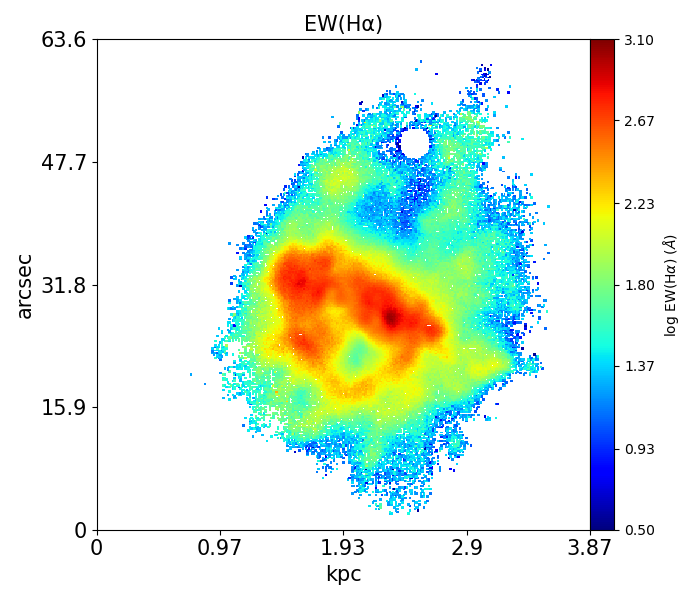}
   \includegraphics[width=6cm]{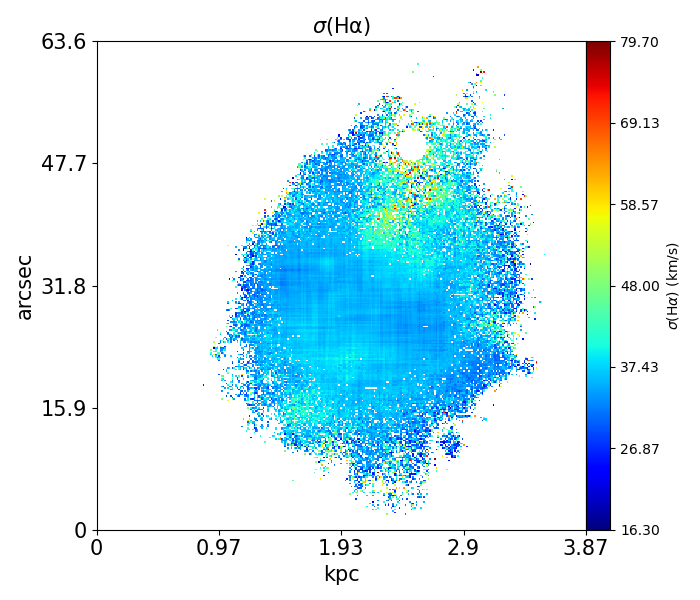}\\
   \includegraphics[width=6cm]{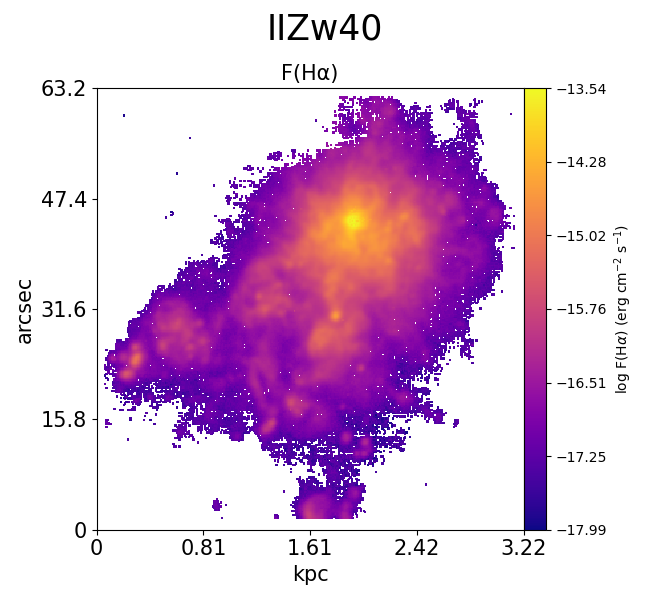}
   \includegraphics[width=6cm]{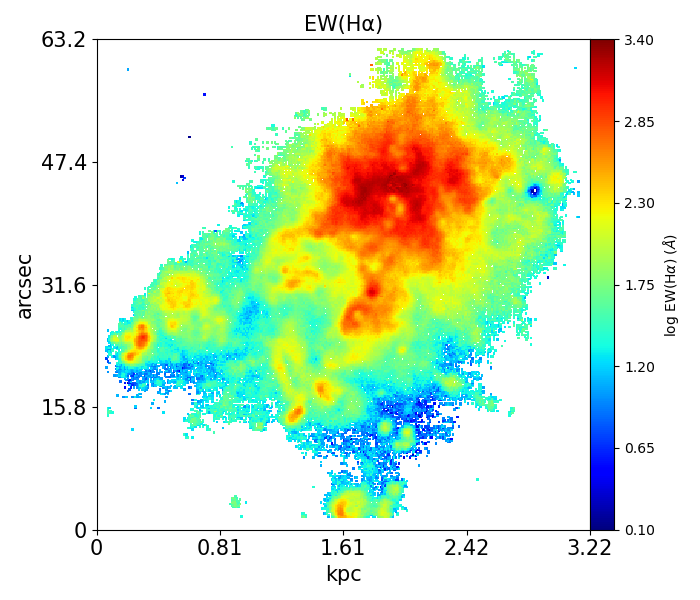}
   \includegraphics[width=6cm]{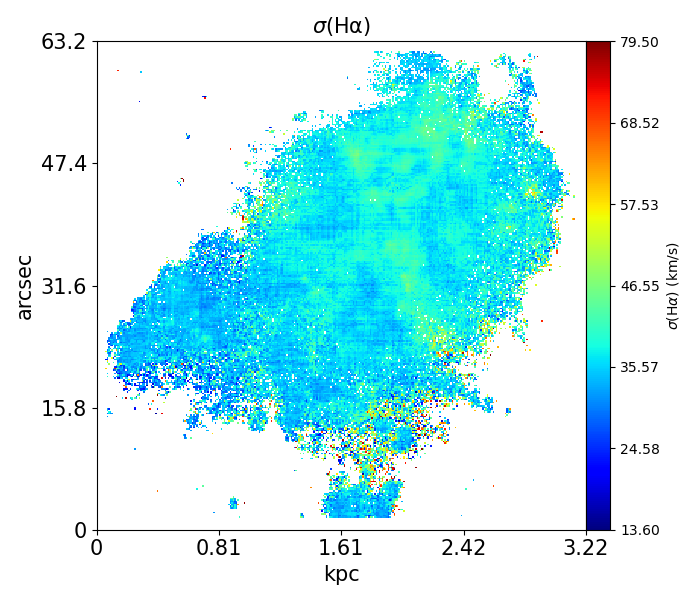}
   \caption{F(H$\alpha$) maps (left panels), EW(H$\alpha$) maps (middle), and
    $\sigma$(H$\alpha$) maps (right). 
   Pixels/spaxels with a S/N < 5 were removed. 
   The colour scale at the right of each panel indicates the F(H$\alpha$),
   EW(H$\alpha$) and $\sigma$(H$\alpha$) levels, respectively. 
   North is up and east is left.}
   \label{fig_Ha}%
\end{figure*}

Figure \ref{fig_Ha} shows our H$\alpha$ flux F(H$\alpha$), H$\alpha$ equivalent
width EW(H$\alpha$), and velocity dispersion $\sigma$(H$\alpha$) maps for UM 462
(upper panels) and IIZw40 (lower panels).
We calculated $\sigma$(H$\alpha$) using the relation between velocity
dispersion and full width at half maximum (FWHM), $\sigma$ = FWHM/2.35, after
correcting the measured FWHM for  the instrumental
broadening (FWHM$_{\mathrm{inst}} \sim 2.6$ \AA\ near H$\alpha$) and thermal
broadening at 10$^4$ K.
Our H$\alpha$ emission and EW(H$\alpha$) maps display similar 
morphologies to the H$\beta$ maps in \cite{Lagos2007}, while the $\sigma$(H$\alpha$)
maps resemble those shown by \cite{MonrealIbero2023} and \cite{Bordalo2009}
for UM 462 and IIZw40, respectively, and they are also consistent with the maps
presented by \cite{Marasco2023}, who used the same dataset.

The F(H$\alpha$) maps for UM 462 and IIZw40 reveal a network
of extended low-luminosity ﬁlamentary structures beyond the main HII regions 
tracing most of the DIG distribution. 
The EW(H$\alpha$) maps offer complementary information
by measuring the relative strength of H$\alpha$ emission compared to the underlying
stellar continuum. High EW values in the  maps indicate
the presence of young, massive star-forming regions, while lower values correspond to
areas dominated by older stellar populations and DIG. In both galaxies, the EW
distribution closely  follows their morphological structure, highlighting the
connection between star formation and the ionised gas environment.
The velocity dispersion maps display a striking pattern, 
with the most kinematically disturbed regions located along ﬁlamentary structures and
shells, or in cavities between these features.
In both galaxies, these regions reach values $\gtrsim$50 km s$^{-1}$.
Those values are consistent with low-velocity shocks and line proﬁle velocities of
active galactic nuclei (AGNs) observed in 
high-redshift galaxies and/or star-forming clumps
\cite[e.g.][]{Epinat2010,Elmegreen2007}.

\subsection{DIG definition and H$\alpha$ surface brightness ($\Sigma$(H$\alpha$))
bins}\label{sec_bins}

\begin{figure}
        \centering
        \includegraphics[width=8.8cm]{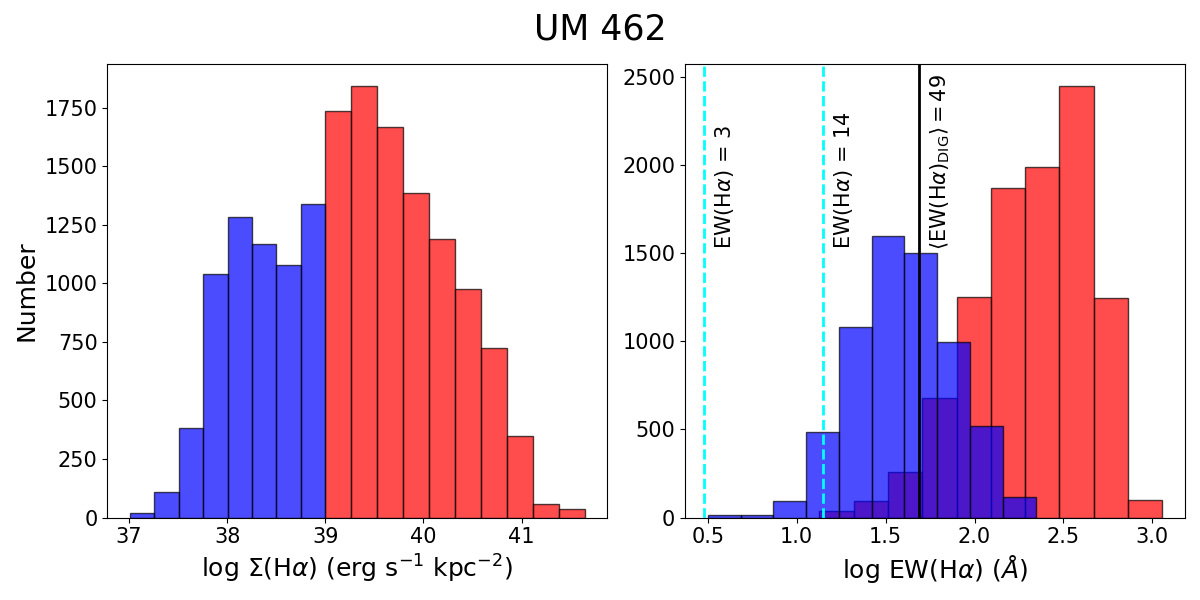}\\
        \includegraphics[width=8.8cm]{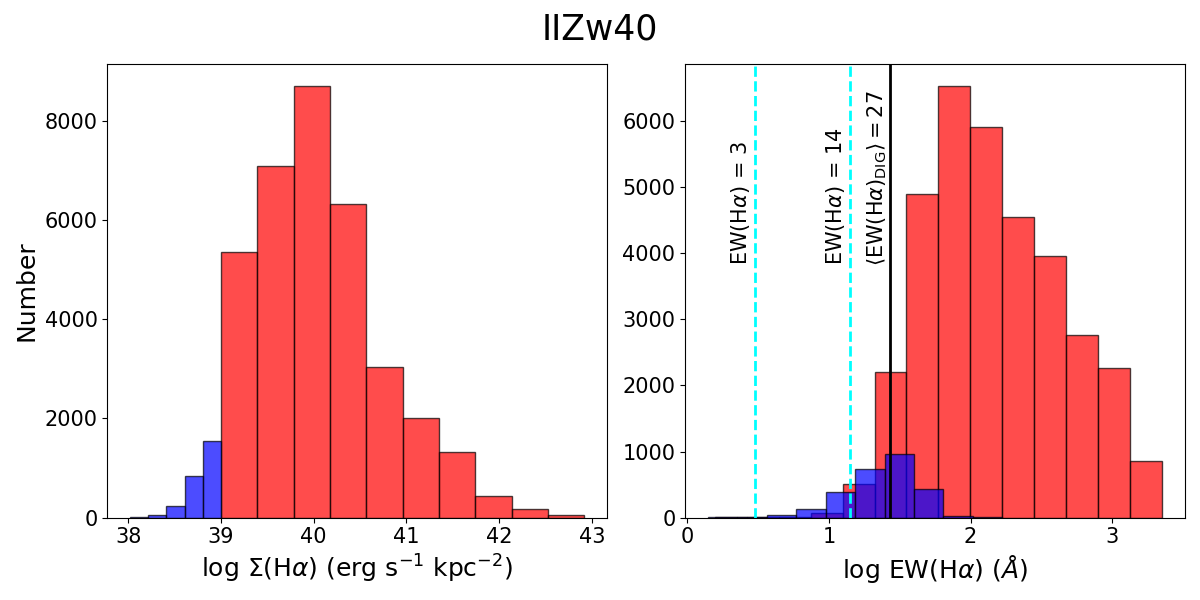}
        \caption{Histograms of $\Sigma$(H$\alpha$) and EW(H$\alpha$).
        The HII-dominated spaxel values are shown in red, while DIG-dominated spaxel
        values are in blue. The dotted cyan lines indicate the classification by
        \cite{Lacerda2018}, where DIG-dominated regions have EW(H$\alpha$) < 3 $\AA$ 
        and HII region-dominated areas have EW(H$\alpha$) > 14 $\AA$. The mean spaxel 
        values of our DIG-dominated areas are indicated by the vertical black lines.}
        \label{Histograms}%
\end{figure}

Diffuse ionised gas is commonly defined using $\Sigma$(H$\alpha$) and/or
EW(H$\alpha$). \cite{Zhang2017} defined the DIG-dominated regions
as being those with $\Sigma$(H$\alpha$) $<$ 10$^{39}$ erg s$^{-1}$ kpc$^{-2}$.
Using this definition, we found that 61\% of the ionised gas is located
in the HII-dominated regions of UM 462, and 39\% is in the DIG-dominated regions. 
For IIZw40, 93\% of the ionised gas in the HII-dominated regions, and 7\% is in the
DIG-dominated regions. 
Using CALIFA data, \cite{Lacerda2018} classified DIG-dominated regions as those with
EW(H$\alpha$) < 3 $\AA$, while HII-dominated regions had EW(H$\alpha$) > 14 $\AA$.
Using this definition, only $\sim$2\% of the data points, for both galaxies, have
EW(H$\alpha$) $<$ 14 $\AA$, implying that several ionisation processes contribute to
the DIG in UM 462 and IIZw40.
The ISM in low-mass galaxies is primarily photoionised by OB stars, but shock 
ionisation from star formation-driven feedback can extend over a large fraction of the
nebular region \citep{Bordalo2009}. Given that EW(H$\alpha$) is highly influenced 
by the local ISM conditions, the DIG will also be affected by the starburst
energy injection. 
In galaxies with strong star formation, the DIG shows large EW(H$\alpha$) values
\citep{Lugo-Aranda2024}. This is exactly what we find in our two studied galaxies. 
Figure \ref{Histograms} shows the histograms of $\Sigma$(H$\alpha$) and
EW(H$\alpha$) for HII (red) and DIG-dominated (blue) spaxels for UM 462 and IIZw40 
using the $\Sigma$(H$\alpha$) definition of DIG.

\begin{figure}
        \centering
        \includegraphics[width=8cm]{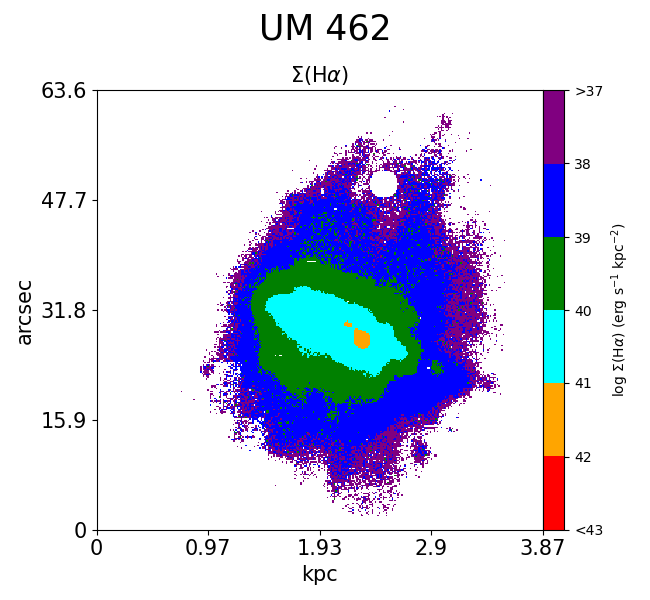}
        \includegraphics[width=8cm]{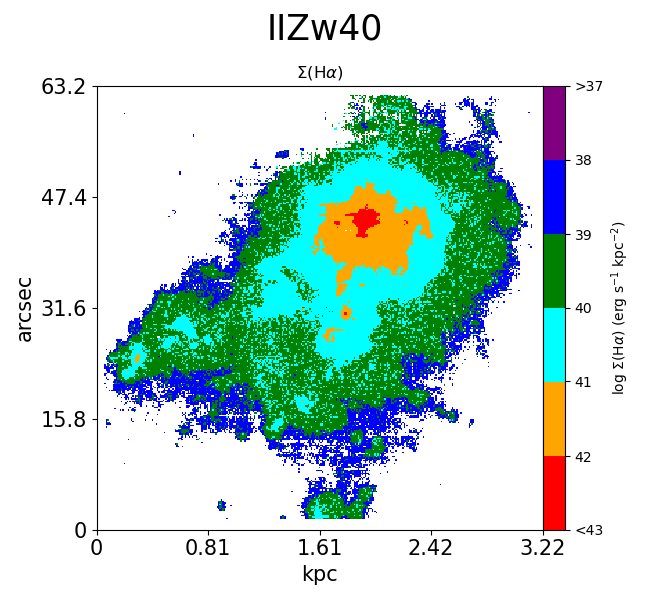}
        \caption{H$\alpha$ Surface brightness maps. The colour-code represents
        the regions of the galaxies divided into different $\Sigma$(H$\alpha$)
        bins of 37--$<$38 (purple), 38--$<$39 (blue), 39--$<$40 (green), 40--$<$41
        (cyan), 41--$<$42 (orange), and 42--$<$43 (red) in units of log erg s$^{-1}$
        kpc$^{-2}$. North is up, and east is left.}
        \label{Regions}%
\end{figure}

In this study, we use the $\Sigma$(H$\alpha$) criterion of \cite{Zhang2017} to
distinguish between HII- and DIG-dominated regions.
To achieve this, we defined several regions within the galaxies grouped into
different log($\Sigma$(H$\alpha$)) bins.
Figure \ref{Regions} shows regions of the galaxies based on their $\Sigma$(H$\alpha$)
bins 37 -- $<$38, 38 -- $<$39, 39 -- $<$40, 40 -- $<$41, 
41 -- $<$42, and 42 -- $<$43 in units of log (erg s$^{-1}$ kpc$^{-2}$). 
This figure highlights the more extended distribution of DIG in UM\,462, which is
reflected in its significantly higher fraction of DIG-dominated spaxels 
(as defined by $\Sigma$(H$\alpha$); see Fig. \ref{Histograms}) compared to IIZw40.

\subsection{Emission line ratios}\label{sec_emi_ratios}

\begin{figure*}
        \centering
        \includegraphics[width=18cm]{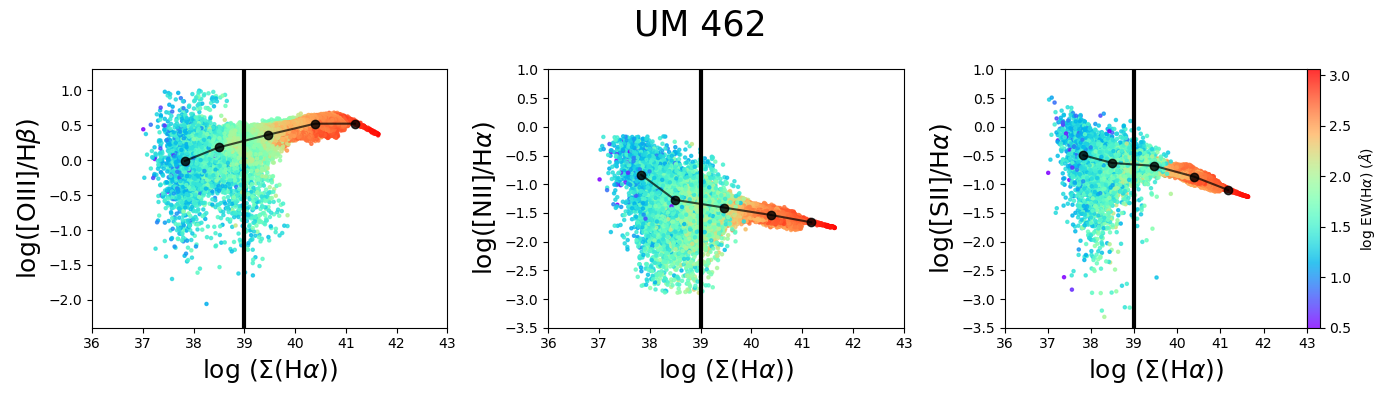}\\
        \includegraphics[width=18cm]{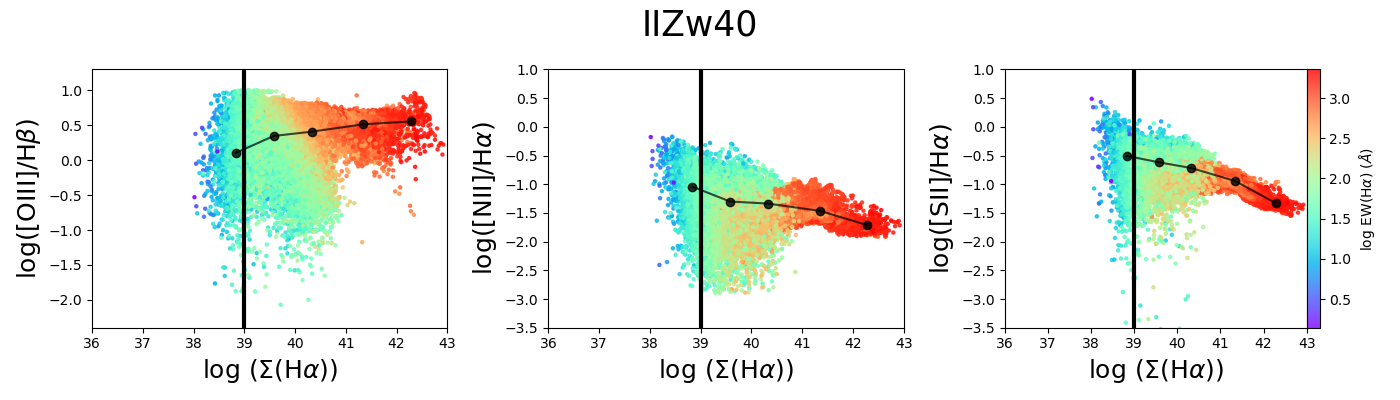}
        \caption{Emission line ratios log([OIII]$\lambda$5007/H$\beta$), log([SII]$\lambda\lambda$6717,6731/H$\alpha$), 
        and log([NII]$\lambda$6584/H$\alpha$) versus
        log($\Sigma$(H$\alpha$)). The vertical line indicates the threshold
        distinguishing DIG-dominated regions, defined as areas with
        $\Sigma$(H$\alpha$) $<$ 10$^{39}$ erg s$^{-1}$ kpc$^{-2}$. The black data
        points correspond to the mean values within the different $\Sigma$(H$\alpha$)
        bins defined in Sect. \ref{sec_bins}.  
        There is a clear increasing pattern of [SII]/H$\alpha$
        and [NII]/H$\alpha$ ratios and a decreasing trend of [OIII]/H$\beta$ as
        $\Sigma$(H$\alpha$) decreases. The colour scale at the right of each panel
        indicates the EW(H$\alpha$).}
        \label{emission_ratios}%
\end{figure*}

\begin{figure}
        \centering
        \includegraphics[width=9cm]{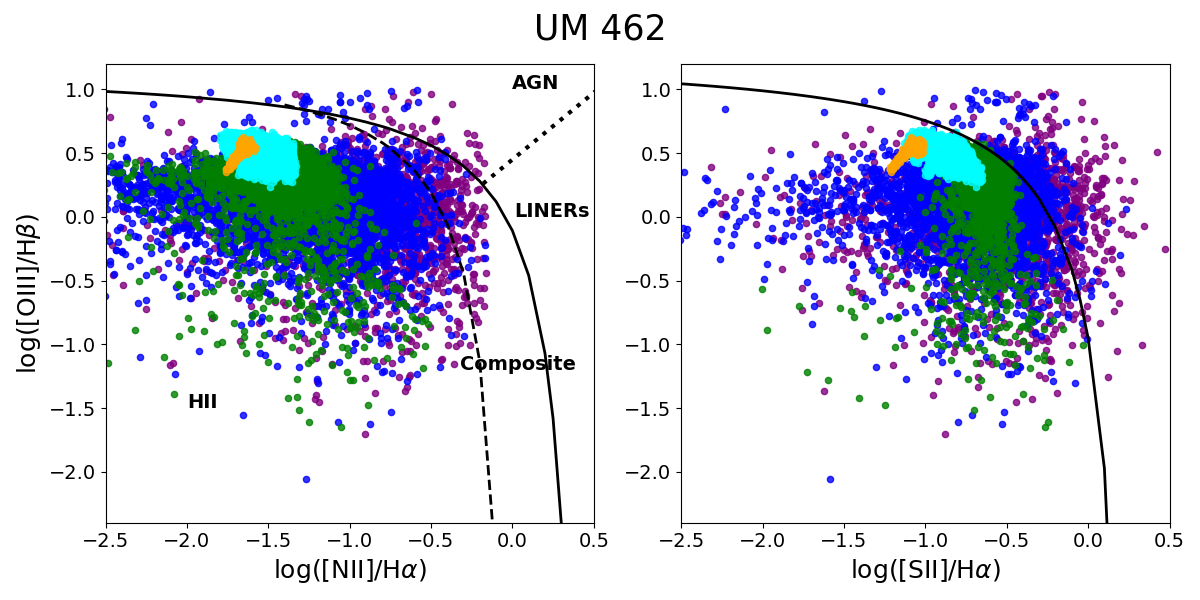}\\
        \includegraphics[width=9cm]{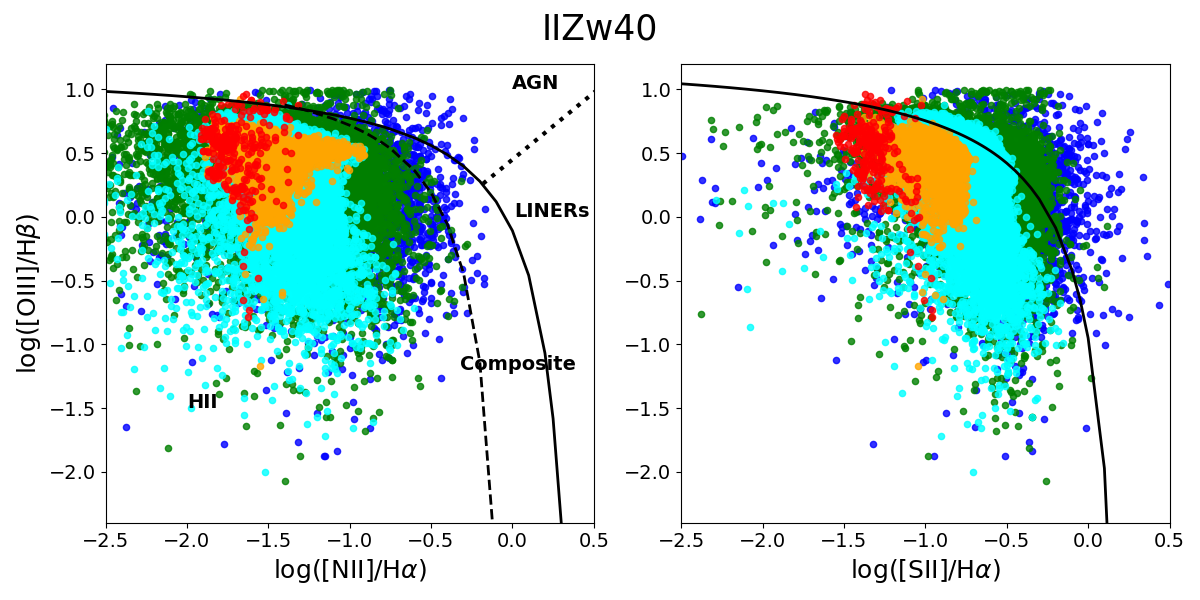}
        \caption{BPT diagrams log([OIII]/H$\beta$ versus log([NII]/H$\alpha$) and log([OIII]/H$\beta$) versus log([SII]/H$\alpha$ for UM 462 (upper panels) 
        and IIZw40 (lower panels). The colour code represents the spaxels of the
        galaxies divided into the different log($\Sigma$(H$\alpha$)) bins of $<$38--37
        (purple), $<$39--38 (blue), $<$40--39 (green), $<$41--40 (cyan), $<$42--41
        (orange), and $<$43--42 (red) in units of log erg s$^{-1}$ kpc$^{-2}$. 
        In the BPT diagrams, we include the \cite{Kewley2001} (solid lines), 
        \cite{Kauffmann2003} (dashed lines), and \cite{Schawinski2007} (dotted lines)
        model boundary lines, which divide regions dominated by star-forming HII,
        composite, and AGN/LINERs.}
        \label{BPT_diagrams}%
\end{figure}

In Fig. \ref{emission_ratios} we show the log([OIII]/H$\beta$),
log([NII]/H$\alpha$), and log([SII]/H$\alpha$) emission line ratios versus
log($\Sigma$(H$\alpha$)) for both galaxies. The vertical lines in this figure
indicate the threshold distinguishing DIG-dominated regions when using the
$\Sigma$(H$\alpha$) definition.
We confirm that log([NII]/H$\alpha$) and log([SII]/H$\alpha$) emission line ratios 
are enhanced in the DIG regions, where we also find the lowest EW(H$\alpha$) values.
In Fig. \ref{BPT_diagrams} we use those emission line ratios in conjunction with
the \cite{Kewley2001} starburst and the \cite{Kauffmann2003} parameters to demarcate
regions of OB-star-driven ionisation, mixed ionisation and AGN, shock or HOLMES
ionisation in the classic Baldwin-Phillips-Terlevich \citep[BPT;][]{Baldwin1981}
diagrams. Our results show that radiation from OB stars  
is likely the predominant ionisation source, with a smaller contribution from other
sources in our galaxies. 
A significant fraction of the data points, mainly in the
log([SII]/H$\alpha$) versus log([OIII]/H$\beta$) diagram, are above the boundary
limits between HII and AGN/low-ionization nuclear emission-line regions (LINERs) in
the BPT diagrams. This is most likely attributable to shock ionisation. 
We discuss this further in Sect. \ref{Sect_discussion}.

\subsection{Abundance determinations}

\begin{figure*}
        \centering
        \includegraphics[scale=0.45]{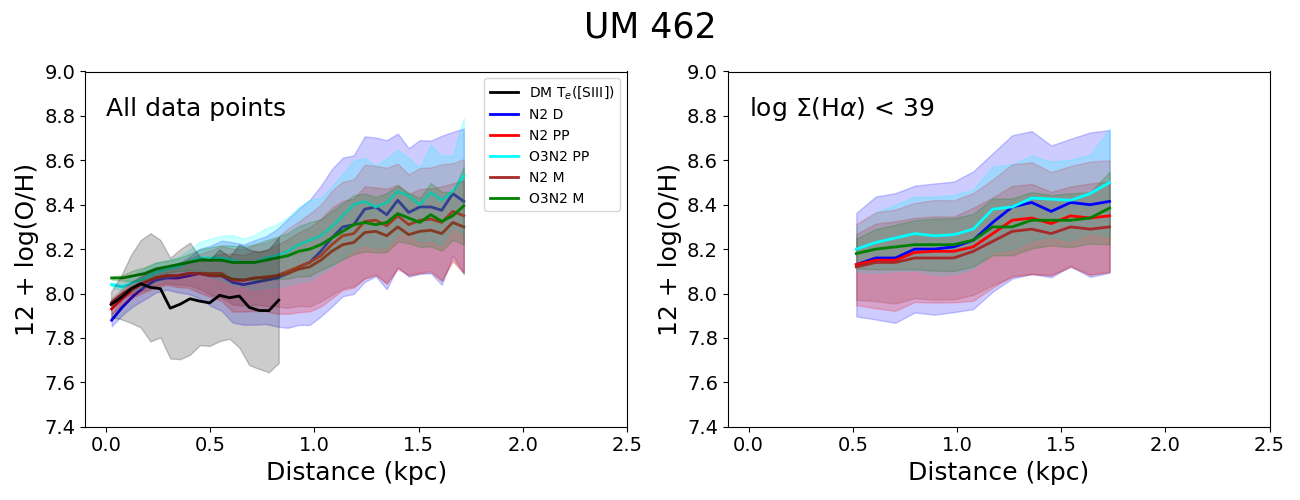}\\
        \includegraphics[scale=0.45]{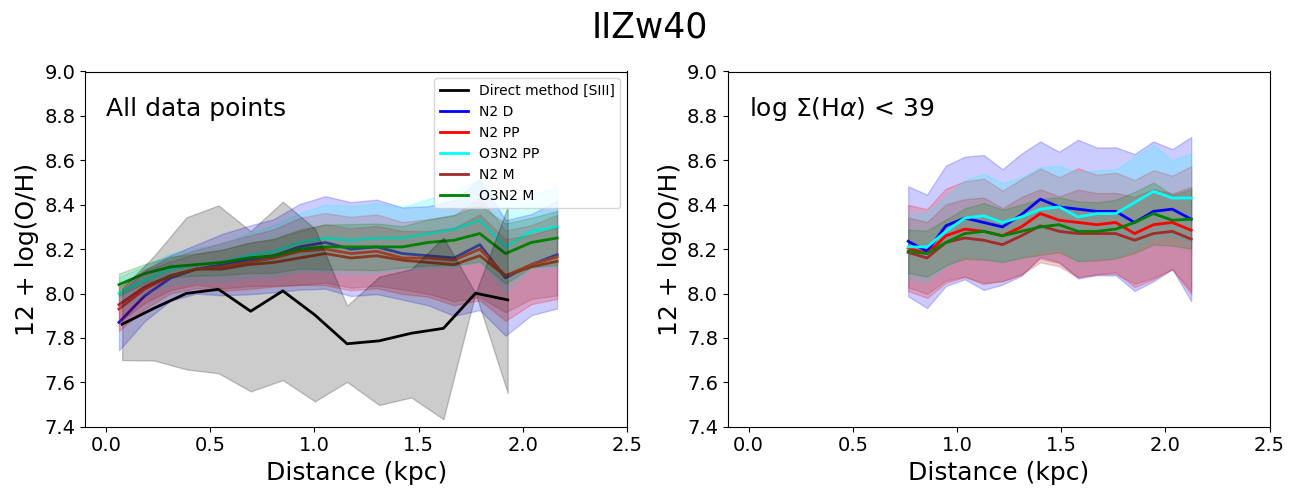}
        \caption{Mean 12 + log(O/H) profiles for all data points (left panels)
        and those with log $\Sigma$(H$\alpha$) $<$ 39 in units of log erg s$^{-1}$
        kpc$^{-2}$ (right panels) derived using the direct method T$_e$([SIII]) and
        the N2 and O3N2 strong-line calibrators from  \cite{Denicolo2002} (D),
        \cite{Pettini2004} (PP), and \cite{Marino2013} (M). The standard deviation of
        the 12 + log(O/H) values are indicated by the shaded regions. A clear
        increase in mean metallicity is observed with distance when using all
        HII-based calibrators as well as the N2 and O3N2 indices for both the full
        dataset and the subset with log $\Sigma$(H$\alpha$) $<$ 39. When applying the
        direct method, the metallicity distribution across the ISM appears flat.
        }
        \label{logOH_all_DIG}%
\end{figure*}

\begin{figure*}[!h]
        \centering
        \includegraphics[width=14cm]{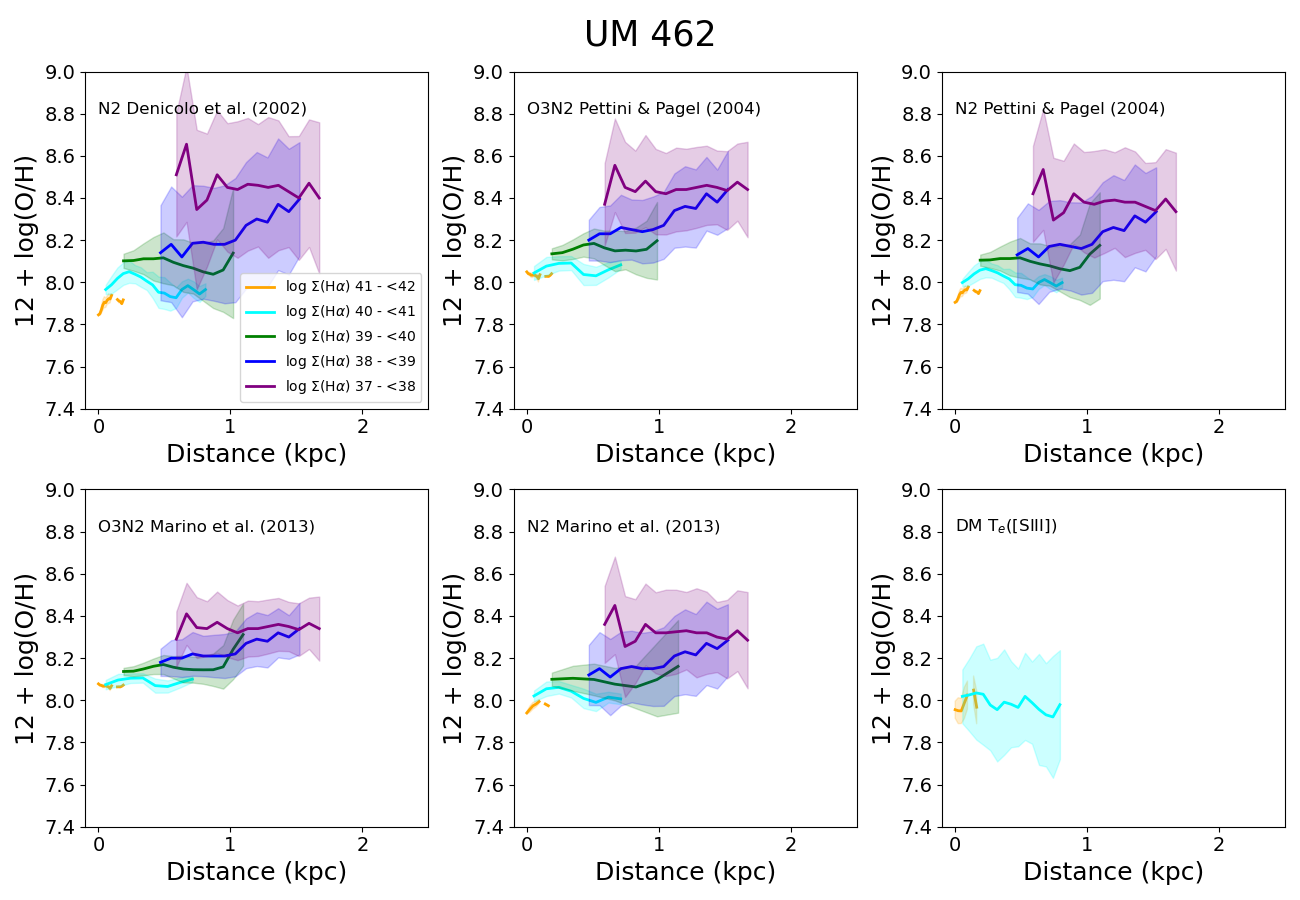}\\
        \includegraphics[width=14cm]{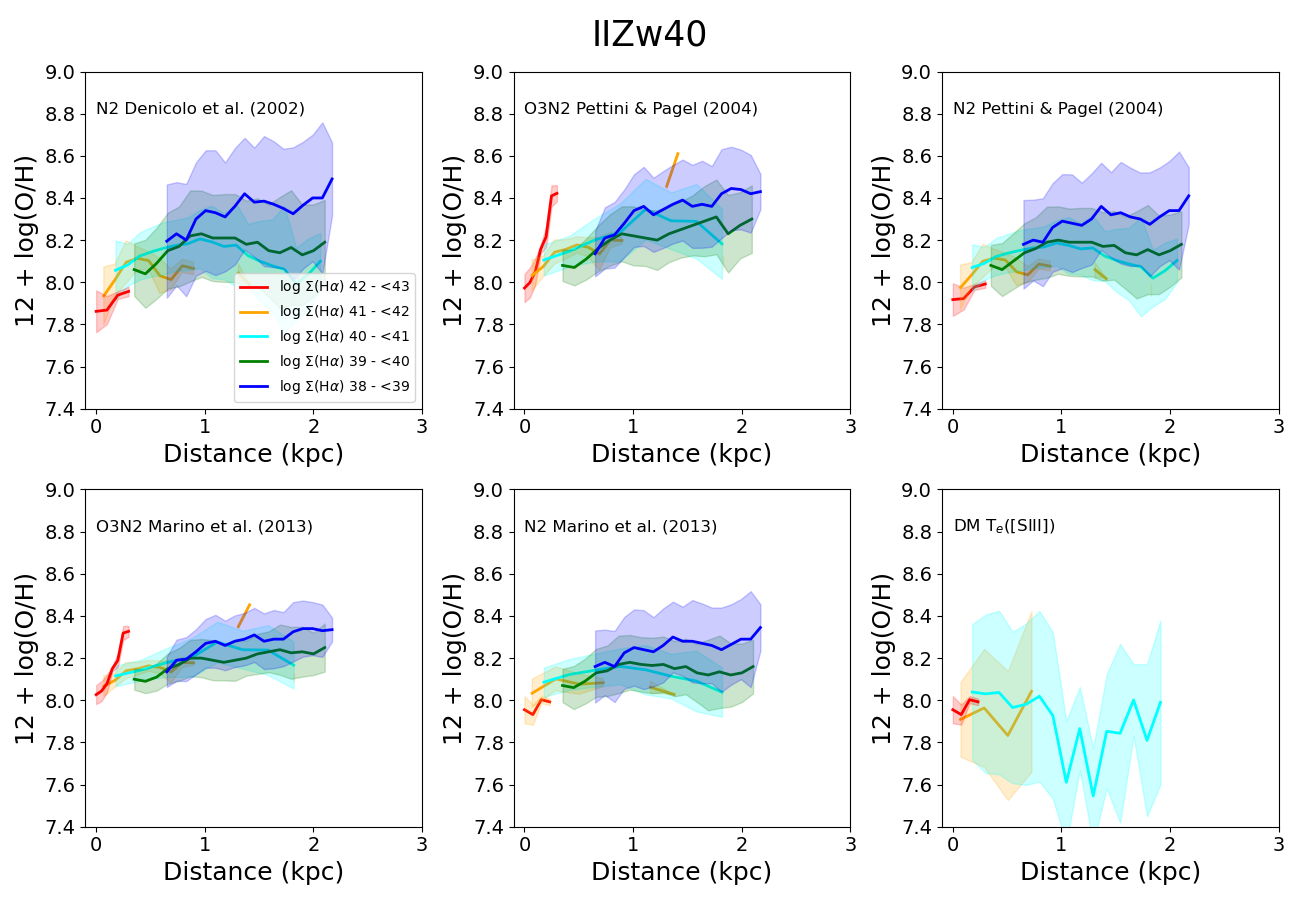}
        \caption{Mean 12 + log(O/H) profiles for levels of $\Sigma$(H$\alpha$) 
        for UM 462 (upper panels) and IIZw40 (lower panels). The standard deviation
        for the different 12 + log(O/H) values within each bin are indicated by the
        shaded regions. A clear increase in the mean metallicity values can be
        observed with decreasing $\Sigma$(H$\alpha$) when using both the N2 and O3N2
        indices across all calibrators. A slight upwards trend in 12+log(O/H) is
        also observed with the direct method in the HII-dominated regions. 
        However, this difference is not statistically significant within the
        uncertainties.
        }
        \label{logOH_S_cal_UM462_IIZW40}%
\end{figure*}

To calculate the electron temperature, density, and oxygen abundances,
in the HII-dominated pixels and/or spaxels 
we used the \textsc{PyNeb} package \citep{Luridiana2015}, version 1.1.13,
with the default set of atomic data (transition probabilities and collision
cross-sections). We used \texttt{S2.getTemDen()}, `L(6717) / L(6731)', and
T$_e$=10$^4$ K for the n$_e$, while the T$_e$([SIII]) was determined using
\texttt{S3.getTemDen()}, `L(6312) / L(9069)`, and the previously determined n$_e$.
Ionic abundances were calculated using the \texttt{Atom.getTemDen()} method
and the emission lines [OIII]$\lambda\lambda$4959,5007 and
[OII]$\lambda\lambda$7320,7331 for the determination of O$^{+}$/H$^{+}$ and
O$^{++}$/H$^{+}$, respectively. We considered the total oxygen abundance as
O/H = O$^{+}$/H$^{+}$ + O$^{++}$/H$^{+}$.
Our spatially resolved results, in the case of UM 462, are similar to those from
\cite{MonrealIbero2023}, indicating we achieved reliable and consistent results for
both galaxies. We found a mean 12 + log(O/H) of 8.01 and 8.0
with a standard deviation of 0.22 dex and 0.32 dex for UM 462 and IIZw40,
respectively. Those results are consistent with the integrated values of 8.02 
and 8.07 found by \cite{MonrealIbero2023} and \cite{Bordalo2011}
for UM 462 and IIZw40, respectively.
Additionally, we determined the oxygen abundances using the indices
N2 = log([NII]$\lambda$6584/H$\alpha$) and O3N2 =
log([OIII]$\lambda$5007/H$\beta \times$ H$\alpha$/[NII]$\lambda$6584), together with
the calibrators from \cite{Denicolo2002} (D), \cite{Pettini2004} (PP), and
\cite{Marino2013} (M).
The O/H abundances in the different $\Sigma$(H$\alpha$) bins were derived using
the HII-based methods described above after removing the data points above the
\cite{Kauffmann2003} demarcation (dashed line in Fig. \ref{BPT_diagrams}) in order 
to exclude regions ionised by sources other than young stars in HII regions.
Figure \ref{logOH_all_DIG} shows the oxygen abundance versus 
distance (in kiloparsecs) from the H$\alpha$ maxima for all spaxels (left panels) 
and those classified as DIG (right panels). Figure \ref{logOH_S_cal_UM462_IIZW40}
presents the same relation, divided into $\Sigma$(H$\alpha$)
bins, with each panel showing results from different calibrators and the direct
method. In Table \ref{Table_2} we present the mean and standard deviation of the
12 + log(O/H) values for all data points as well as for those with $\Sigma$(H$\alpha$)
< 39 (DIG-dominated).
In Table \ref{Table_3}, we show the same quantities but
computed across the $\Sigma$(H$\alpha$) bins defined in this study. 
This table also presents the EW(H$\alpha$), the ionisation parameter 
(discussed in Sect. \ref{Sect_discussion}), and the $\sigma$(H$\alpha$) values for
the different $\Sigma$(H$\alpha$) bins.
We observed a clear increase in the mean metallicity values
with decreasing $\Sigma$(H$\alpha$) when using both the N2 and O3N2
indices and all calibrators (see Figs. \ref{logOH_all_DIG} and
\ref{logOH_S_cal_UM462_IIZW40}). 
An increasing trend in 12 + log(O/H) was also observed when using the direct method in
the HII-dominated regions. However, the difference is not statistically significant
within the uncertainties.

\begin{table}[ht]
        \caption{Mean and standard deviation of 12 + log(O/H) for 
        all and $\Sigma$(H$\alpha$) < 39 (DIG-dominated) data points. 
        Values were derived using the direct method (DM)
        T$_e$([SIII]) and the N2 and O3N2 calibrators from \citet{Denicolo2002} (D),
        \citet{Pettini2004} (PP), and \citet{Marino2013} (M).}
        \centering
        \begin{tabular}{lcc}
                \hline\hline
                & \multicolumn{2}{c}{12 + log(O/H)} \\
                Method & UM 462 & IIZw 40 \\
                \hline
                \multicolumn{3}{l}{All data points} \\
                DM T$_e$([SIII]) & 8.01 $\pm$ 0.22 & 8.00 $\pm$ 0.33 \\
                N2 D             & 8.12 $\pm$ 0.25 & 8.14 $\pm$ 0.20 \\
                O3N2 PP          & 8.20 $\pm$ 0.17 & 8.20 $\pm$ 0.15 \\
                N2 PP            & 8.12 $\pm$ 0.19 & 8.14 $\pm$ 0.15 \\
                O3N2 M           & 8.18 $\pm$ 0.11 & 8.18 $\pm$ 0.10 \\
                N2 M             & 8.11 $\pm$ 0.16 & 8.13 $\pm$ 0.13 \\
                \hline
                \multicolumn{3}{l}{Data points with log $\Sigma$(H$\alpha$) $<$ 39 (DIG)}\\
                DM T$_e$([SIII]) & ... & ... \\
                N2 D             & 8.24 $\pm$ 0.33 & 8.34 $\pm$ 0.29 \\
                O3N2 PP          & 8.31 $\pm$ 0.19 & 8.31 $\pm$ 0.21 \\
                N2 PP            & 8.21 $\pm$ 0.26 & 8.29 $\pm$ 0.23 \\
                O3N2 M           & 8.25 $\pm$ 0.13 & 8.25 $\pm$ 0.14 \\
                N2 M             & 8.18 $\pm$ 0.21 & 8.25 $\pm$ 0.19 \\
                \hline
        \end{tabular}
        \tablefoot{Values are given as mean $\pm$  standard deviation.}
                \label{Table_2}
\end{table}

\begin{table*}[ht]
        \caption{Mean and standard deviation of 12+log(O/H) for the $\Sigma$(H$\alpha$) bins. The values were derived
        using the direct method (DM) T$_e$([SIII]) and the N2 and O3N2 calibrators
        from \citet{Denicolo2002} (D), \citet{Pettini2004} (PP), and
        \citet{Marino2013} (M).}
        \centering
        \begin{tabular}{lcccccc}
                \hline
                & \multicolumn{6}{c}{12 + log(O/H)} \\
                log($\Sigma$(H$\alpha$)) & 37--<38 & 38--<39 & 39--<40 & 40--<41 & 41--<42 & 42--<43 \\
                \hline
                \textit{UM 462} \\
                DM T$_e$([SIII]) & ... & ... & ... & 8.02 $\pm$ 0.22 & 7.98 $\pm$ 0.08 & ... \\
                N2 D      &  8.39 $\pm$ 0.31 &  8.17 $\pm$ 0.29 & 8.09 $\pm$ 0.13 & 8.00 $\pm$ 0.06 & 7.90 $\pm$ 0.03 & ... \\
                O3N2 PP   &  8.44 $\pm$ 0.20 &  8.27 $\pm$ 0.16 & 8.18 $\pm$ 0.10 & 8.07 $\pm$ 0.04 & 8.03 $\pm$ 0.01 & ... \\
                N2 PP     &  8.33 $\pm$ 0.25 &  8.16 $\pm$ 0.22 & 8.09 $\pm$ 0.10 & 8.02 $\pm$ 0.05 & 7.95 $\pm$ 0.02 & ... \\
                O3N2 M    &  8.34 $\pm$ 0.13 &  8.23 $\pm$ 0.11 & 8.16 $\pm$ 0.07 & 8.09 $\pm$ 0.03 & 8.07$\pm$0.01 & ... \\
                N2 M      &  8.28 $\pm$ 0.20 &  8.14 $\pm$ 0.18 & 8.09 $\pm$ 0.10 & 8.02 $\pm$ 0.05 & 7.95 $\pm$ 0.02 & ... \\               
                \hline
                log(EW(H$\alpha$))       & 1.37 $\pm$ 0.22 & 1.67 $\pm$ 0.25 & 2.18 $\pm$ 0.29 & 2.57 $\pm$ 0.16 & 2.88 $\pm$ 0.09 & ... \\
                log(U)                   & -3.18 $\pm$ 0.77 & -2.98 $\pm$ 0.54 & -3.0 $\pm$ 0.16 & -2.68 $\pm$ 0.16 & -2.3 $\pm$ 0.10 & ... \\
                log($\sigma$(H$\alpha$)) & 1.55 $\pm$ 1.00 & 1.57 $\pm$ 0.73 & 1.56 $\pm$ 0.32 & 1.55 $\pm$ 0.03 & 1.54 $\pm$ 0.00 & ... \\
                \hline
                \textit{IIZw 40} \\
                DM T$_e$([SIII]) & ... & ... & ... & 8.08 $\pm$ 0.38 & 7.94 $\pm$ 0.24 & 7.80 $\pm$ 0.21 \\
                N2 D      & ... & 8.29 $\pm$ 0.29 & 8.15 $\pm$ 0.21 & 8.13 $\pm$ 0.14 & 8.05 $\pm$ 0.11 & 7.88 $\pm$ 0.09 \\
                O3N2 PP   & ... & 8.34 $\pm$ 0.19 & 8.21 $\pm$ 0.15 & 8.20 $\pm$ 0.13 & 8.11 $\pm$ 0.07 & 8.03 $\pm$ 0.10 \\
                N2 PP     & ... & 8.26 $\pm$ 0.23 & 8.14 $\pm$ 0.17 & 8.12 $\pm$ 0.11 & 8.07 $\pm$ 0.09 & 7.93 $\pm$ 0.07 \\
                O3N2 M    & ... & 8.27 $\pm$ 0.13 & 8.18 $\pm$ 0.10 & 8.18 $\pm$ 0.09 & 8.12 $\pm$ 0.05 & 8.06 $\pm$ 0.06 \\
                N2 M      & ... & 8.22 $\pm$ 0.18 & 8.13 $\pm$ 0.13 & 8.12 $\pm$ 0.11 & 8.07 $\pm$ 0.09 & 7.93 $\pm$ 0.07 \\
                \hline
                log(EW(H$\alpha$))       & ... & 1.37 $\pm$ 0.24 & 1.86 $\pm$ 0.30 & 2.34 $\pm$ 0.36 & 2.96 $\pm$ 0.18 & 3.14 $\pm$ 0.18 \\
                log(U)                   & ... & -3.2 $\pm$ 0.61 & -3.06 $\pm$ 0.35 & -2.92 $\pm$ 0.24 & -2.52 $\pm$ 0.22 & -1.94 $\pm$ 0.19 \\
                log($\sigma$(H$\alpha$)) & ... & 1.57 $\pm$ 0.98 & 1.56 $\pm$ 0.56 & 1.57 $\pm$ 0.43 & 1.57 $\pm$ 0.32 & 1.58 $\pm$ 0.30 \\
                \hline
        \end{tabular}
        \tablefoot{Values are given as mean $\pm$  standard deviation. 
        Mean values of log(EW(H$\alpha$)) ($\AA$), log($U$) and
        log($\sigma$(H$\alpha$)) (km s$^{-1}$) are also reported in the corresponding
        $\Sigma$(H$\alpha$) bins.}
        \label{Table_3}
\end{table*}

\section{Discussion and conclusions}\label{Sect_discussion}

\begin{figure}
        \centering
        \includegraphics[width=9cm]{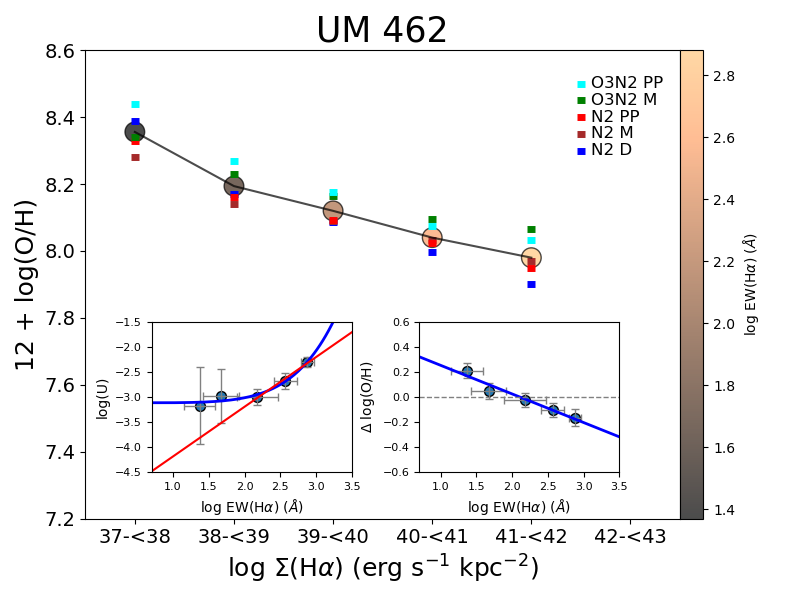}
        \includegraphics[width=9cm]{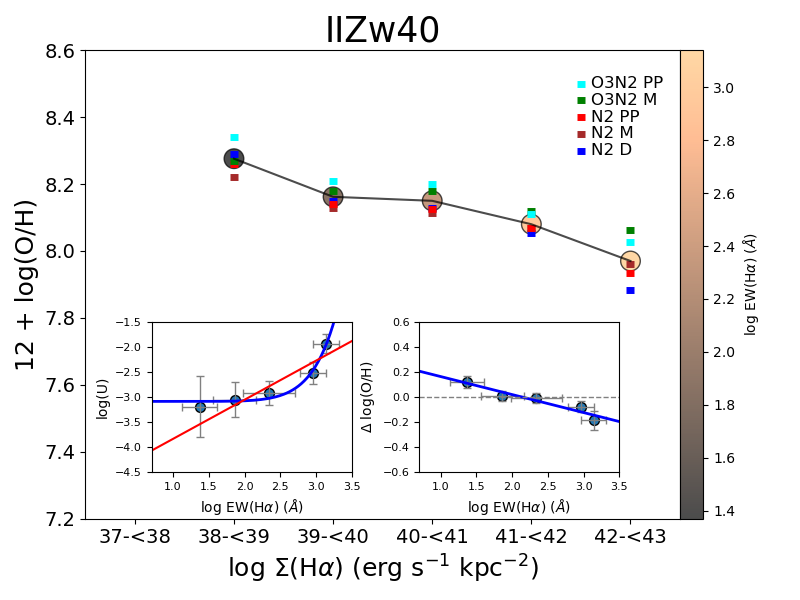}
        \caption{Relationship between the mean 12+log(O/H)
                and surface brightness bins for UM 462 and IIZw40. The values from the
                calibrators used in this study are indicated by the coloured squares
                in the figure. The inset panels show the log(U) (left panels) and the
                difference in O/H relative to the mean value of the galaxies
                ($\Delta$(O/H); right panels) as functions of the EW(H$\alpha$). 
                Oxygen abundances show an inverse correlation with $\Sigma$(H$\alpha$). The DIG-dominated regions exhibit the highest
                oxygen abundances together with the lowest ionisation parameters and
                EW(H$\alpha$) values. We found a linear correlation between
                $\Delta$(O/H) and EW(H$\alpha$).
                }
        \label{FigGam}%
\end{figure}

We now examine how the oxygen abundance and related ionisation properties vary with
H$\alpha$ surface brightness across the galaxies.
In Fig. \ref{FigGam} we present, for each galaxy, the mean 12 + log(O/H) derived
from the calibrators across the $\Sigma$(H$\alpha$) bins.
The colour scale on the right gives the mean EW(H$\alpha$) of each bin.
The left inset panels show the ionisation parameter U (the relative number of ionising
photons compared to the number of hydrogen atoms) as a function of EW(H$\alpha$).
The right inset panels show the deviation of the oxygen
abundance ($\Delta$(O/H)) derived using the calibrators from the mean O/H value as a
function of EW(H$\alpha$) for the same $\Sigma$(H$\alpha$) bins.
This figure reveals a notable correlation between high
metallicity and regions of low $\Sigma$(H$\alpha$).

Historically, the mean 12 + log(O/H) value has been considered as a reliable
representation of the overall ISM chemical composition.
For most HII or BCD galaxies reported in the literature -- including those in this
study -- the evidence \citep[see][]{Lagos2009,Lagos2012,LagosPapaderos2013} suggests
that they are chemically homogeneous when using the HII-based methods.
Using the strong-line calibrators (see Fig. \ref{FigGam}), regions with the lowest
$\Sigma$(H$\alpha$) exhibit 12 + log(O/H) values higher than the mean by $\sim$0.2
dex in UM 462 and $\sim$0.1 dex in IIZw40. In contrast, metallicity differences reach 
about 0.4 dex and 0.3 dex between the highest and lowest $\Sigma$(H$\alpha$) bins 
in UM 462 and IIZw40, respectively.
These results indicate that significant O/H variations are observed in the
DIG-dominated regions (see Figs. \ref{logOH_all_DIG} and
\ref{logOH_S_cal_UM462_IIZW40}) compared to the HII-dominated ones, based on both the direct method
and strong-line calibrators, although the overall chemical composition appears
homogeneous within the uncertainties.

Despite the scatter in the outer low surface brightness regions in
Fig. \ref{BPT_diagrams}, there is a clear increase in the emission line ratios 
[NII]/H$\alpha$ and [SII]/H$\alpha$, and consequently in metallicity, from the centre
outwards, indicating a physically meaningful gradient.
The metallicity derived using N2 and/or O3N2 at large radii 
($\Sigma$(H$\alpha$) bins 37 - $<$38 and 38 - $<$39 in units of 
log erg s$^{-1}$ kpc$^{-2}$) are ionised by several different sources 
(e.g. photon leaked from the HII regions and/or HOLMES) rather than young massive
stars within HII regions. 
According to \cite{Belfiore2022}, the increase in [NII]/H$\alpha$ and
[SII]/H$\alpha$ ratios along with a decrease in [OIII]/H$\beta$ ratios can be
explained by DIG ionisation  driven primarily by leaking radiation from HII regions,
with a minor contribution from HOLMES. 
This may account for the observed increase in emission line ratios as a function
of $\Sigma$(H$\alpha$) in Fig. \ref{emission_ratios}.

Using the [SII]$\lambda\lambda$6717,6731/H$\alpha$ emission line ratios 
in Sect. \ref{sec_emi_ratios} and the parametrisation by \cite{Dors2011}, 
we calculated the ionisation parameter U (see insets in Fig. \ref{FigGam}).
In Table \ref{Table_3} we show the mean and  standard deviation of log(U) values
across the different $\Sigma$(H$\alpha$) bins.
In both galaxies, the metallicity (from the calibrators) relative to the mean
shows a linear corelation with log EW(H$\alpha$) (see Fig. \ref{FigGam}, 
blue lines, right insets). 
Moreover, log(U) in the HII-dominated regions is also correlated with
EW(H$\alpha$) (Fig. \ref{FigGam} left insets, red lines) at least for UM 462. 
A similar relation between $\log(U)$ and $\log(\text{EW(H}\alpha\text){)}$ 
was seen in \cite{Mingozzi2020}.
However, when both HII and DIG-dominated regions are included,
the relation follows a power-law (Fig. \ref{FigGam}, left insets, blue lines)
of the form $\log(U) \propto \log(\text{EW(H}\alpha\text{)})^{\alpha}$, where $\alpha$
should be a characteristic constant of each galaxy. For UM 462 and IIZw40 $\alpha$ =
5.76 and 9.89, respectively. These relationships reveal a continuous and smooth 
variation in the ionisation structure from the central HII regions into the 
low surface brightness regions, with a declining EW(H$\alpha$) and an increasing O/H 
(see Fig. \ref{FigGam}), when using the N2- and O3N2-based calibrators. 
This suggests that the DIG emission is primarily maintained by ionising
photons escaping from the HII regions \citep{McClymont2024} through a porous or 
leaky ISM.

\begin{figure}[!h]
        \centering
        \includegraphics[width=8cm]{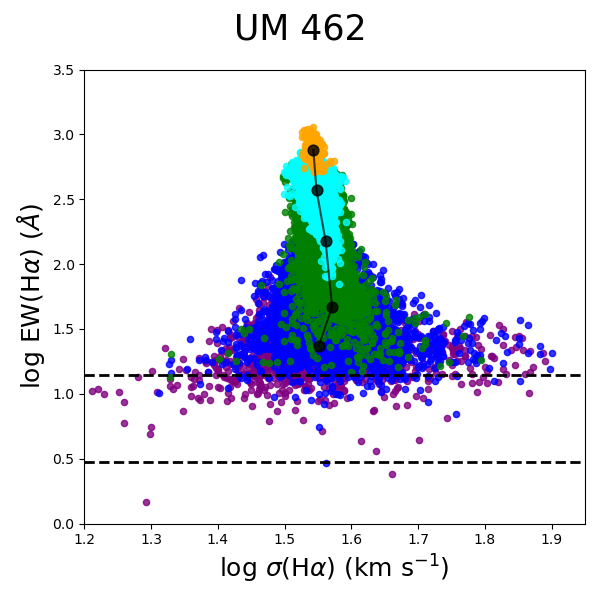}
        \includegraphics[width=8cm]{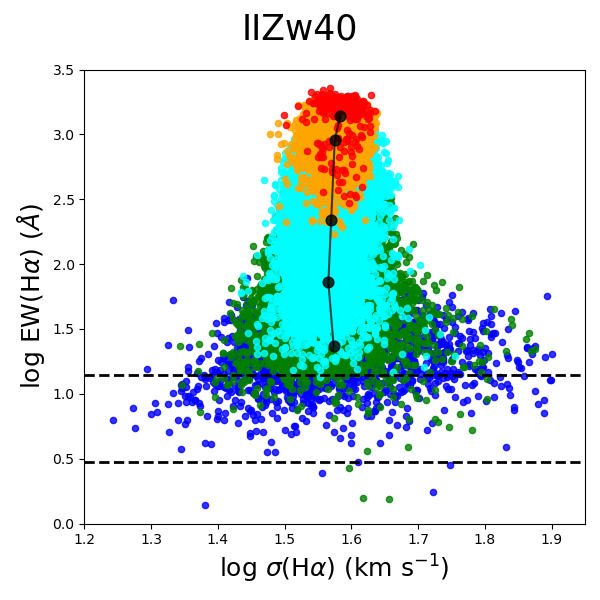}
        \caption{
           Resolved log(EW(H$\beta$)) versus log($\sigma$(H$\alpha$)) for UM 462
           (upper panel) and IIZw40 (lower panel). The colour code represents the
           regions of the galaxies divided into different $\Sigma$(H$\alpha$) bins of
           37-- $<$38 (purple), 38--$<$39 (blue), 39--$<$40 (green), 40--$<$41
           (cyan), 41--$<$42 (orange), and 42--$<$43 (red) in units of log erg
           s$^{-1}$ kpc$^{-2}$. 
           The dotted horizontal lines indicate the classification proposed by
           \cite{Lacerda2018}, where HOLMES DIG-dominated regions have 
           EW(H$\alpha$) < 3 $\AA$ (lower dashed line), and HII-dominated areas have
           EW(H$\alpha$) > 14 $\AA$ (upper dashed line).
           The black data points correspond to the mean values within the
           different $\Sigma$(H$\alpha$) bins.
           The central and vertical bands clearly correspond to the HII-dominated
           regions, whereas the more disturbed, extended bands are those associated
           with the DIG-dominated areas.}
        \label{sigma_Ha}%
\end{figure}

In Fig. \ref{sigma_Ha} for each galaxy we show log(EW(H$\alpha$)) versus
log($\sigma$(H$\alpha$)) across the $\Sigma$(H$\alpha$) bins.   
\cite{Sanchez2024} used this diagram (WHaD diagram) to
separate galaxies and/or regions dominated by star-forming galaxies, retired 
galaxies (HOLMES or post-AGBs), AGNs and weak AGNs.
The shape of this diagram resembles the intensity-velocity
dispersion (I-$\sigma$) relation of emission lines, which is commonly used to study
giant HII regions and HII or BCD galaxies 
\citep[and references therein]{Bordalo2009, MartinezDelgado2007,Moiseev2012}. 
This approach, represented in the \cite{Sanchez2024} diagram,
is well suited to identifying the kinematic features of our studied galaxies.
In our case, most data points are clearly located in the 
HII-dominated areas, with EW(H$\alpha$) > 14 $\AA$ \citep{Lacerda2018}.
The central and vertical bands in Fig. \ref{sigma_Ha} correspond to the
HII-dominated regions, with the bulk of $\sigma$(H$\alpha$) values ranging from
$\sim$30 to $\sim$44 km s$^{-1}$ in UM 462 and from $\sim$25 to $\sim$52 km s$^{-1}$
in IIZw40. In contrast, the more extended and disturbed bands, with
$\sigma$(H$\alpha$) values ranging from $\sim$16 to $\sim$80 km s$^{-1}$ in UM 462
and from $\sim$20 to $\sim$80 km s$^{-1}$ in IIZw40, are associated with the
DIG-dominated regions.
These DIG regions exhibit broader and more asymmetric line profiles
\citep[see][]{Bordalo2009,MonrealIbero2023}, produced by the impact of stellar
mechanical energy on the ISM due to multiple massive star-formation regions, which
are often observed in BCD galaxies \cite[e.g.][]{Lagos2011}.
In our systems, stellar feedback is manifested in large-scale 
super-bubbles and/or shells \citep{MonrealIbero2023,Bordalo2009}. 
For instance, in UM 462, the horn-like DIG structures in the
northern part of the galaxy are interpreted by \citet{MonrealIbero2023} as 
fragmented walls of a super-bubble, which may provide channels for ionising
photons to escape.
In any case, $\sigma$(H$\alpha$) values exceeding $\sim$57 km s$^{-1}$  
($\sim$1.76 in log; the boundary in the WHaD diagram  
that separates HII from AGN-like regions) indicate clear shock contamination in 
some parts of the DIG  \cite[also see Fig. 5 in][]{Marasco2023}.
Consequently, such contamination may affect the reliability of the derived oxygen
abundances in those regions.

High EW(H$\alpha$) values are typically associated with intense 
recent star-formation episodes, as they trace regions dominated by young massive
stars. However, our DIG-dominated regions show relatively high EW(H$\alpha$) values,
although still lower than those found in the HII-dominated regions, with a mean of
49 $\AA$ and 27 $\AA$ for UM 462 and IIZw40, respectively. 
As discussed by \citet{Sanchez2020} and \citet{Lugo-Aranda2024}, the dominant ionising
source of the DIG varies with galaxy morphological type. 
In early-type galaxies, a few HII regions together with older stellar
populations provide potential DIG ionising sources, including HOLMES or 
post-AGB stars, and low-velocity shocks, with the latter typically playing a 
minor role. In contrast, in later-type galaxies, ionisation by leaking photons from 
a large number of HII regions may account for a large fraction of the DIG
emission. Consequently, in areas near HII regions, the DIG is most likely dominated by
leaked radiation and shocks \cite[e.g.][]{GonzalezDiaz2024}, resulting in a more
kinematically disturbed ISM \citep{Levy2019}. 
The EW(H$\alpha$) in these regions is expected to be higher than in those dominated by
HOLMES-like ionising sources \citep{Lugo-Aranda2024}. 
This is exactly what our results indicate.

Regardless of the main mechanism responsible for ionising the DIG, it is important 
to bare in mind that strong-line metallicity calibrators are typically obtained from 
HII-regions, which introduce biases when considering emission from the DIG.
The metallicity gradients observed in star-forming galaxies are interpreted as
signatures of several chemical enrichment processes, including stellar feedback, 
galactic outflows and fountains, inflows of metal-poor gas, mergers, and tidal
interactions. As a result of these mechanisms, star-forming galaxies 
with stellar masses of M$_{\star}$ > 10$^{9.5}$M$_{\odot}$ typically exhibit a
well-defined oxygen abundance gradient of about -0.1 dex/Re across their discs, 
with local deviations such as central drops and outer flattening linked to gas 
flows and mixing \citep{Sanchez2014,Sanchez2020}. In contrast, low-mass systems, 
(M$_{\star}$ < 10$^{9}$M$_{\odot}$) including HII or BCD galaxies, 
are largely chemically homogeneous on galactic scales.
The fact that we find chemical homogeneity in HII-dominated regions using the direct method
suggests that metals in the ISM are distributed `globally' throughout the ISM
of the galaxies on a relatively short timescale \citep{Lagos2016}. 
Moreover, the strong-line N2- and O3N2-based calibrators in Figs.
\ref{logOH_all_DIG} and \ref{logOH_S_cal_UM462_IIZW40} show an increasing abundance
pattern that is more pronounced in DIG-dominated areas.
These results can be extrapolated to galaxies with extended DIG tails, such as tadpole
or cometary-like systems, where previous studies \citep[e.g.][]{SanchezAlmeida2015}
have suggested the presence of metallicity gradients. Such variations, however, can be
misinterpreted as evidence of the infall of metal-poor gas. Because DIG regions are
characterised by physical conditions that differ from those in HII regions, such as
different ionisation sources, lower ionisation parameters, and a more extended spatial
distribution, they can produce artificially inverted metallicity gradients if not
properly accounted for.

Although derivations using HII-based methods may help estimate the level of DIG
contamination, they do not trace real changes in metallicity but rather apparent
variations caused by differing ionisation sources. Again, it is also important to note
that the N2 and O3N2 calibrations are tuned for classical HII regions, not for DIG. 
The N2 index depends on the [NII]/H$\alpha$ ratio, and since nitrogen is a secondary
nucleosynthesis product, its abundance can vary with the star-formation history
(i.e. with N/O; \citealt{Izotov1999,Izotov2006}).
Moreover, both N2 and O3N2 are sensitive to changes in $U$. 
These limitations introduce unquantified uncertainties into the separation of HII
and DIG regions and their interpretation. In BCD galaxies with complex
morphologies (featuring optically thin media that allow enhanced LyC photon leakage and significant DIG fractions), the interpretation of metallicity gradients based on
strong-line calibrators must account for these structural effects to avoid misleading
conclusions about chemical evolution. If our interpretation is correct, similar issues
should also be observable in other local low-mass systems, such as HII or BCD 
galaxies and XMPs as well as in Green Pea galaxies at intermediate redshift and in
high-redshift star-forming galaxies and clumps.

According to \cite{Mannucci2021}, the differences between the spectra of local HII
regions and more distant galaxies are not caused by contamination from the DIG but
by an aperture effect. However, we expect aperture effects to further amplify
this bias, as integrated spectra based on larger apertures tend to include a greater
fraction of DIG emission, which is preferentially located in the outer
parts of the galaxies (see Fig. \ref{Regions}). 
In strong starbursts such as IIZw40, the DIG
emission is difficult to separate from the HII-dominated regions, making DIG
contamination harder to resolve. The difference in the distribution and amount of DIG
in the two galaxies appears to be influenced by their distinct evolutionary states: 
IIZw40 is in the midst of a merger, whereas UM 462, although part of a small group
together with UM 461 \citep{Lagos2018}, presents as a relatively undisturbed galaxy. 
These findings indicate that the interplay between star formation, gas dynamics, and environment
play a crucial role in shaping the properties of the diffuse ionised medium.
Therefore, we conclude that the most likely mechanism for ionising the DIG 
in our sample of local analogues of high-redshift star-forming galaxies 
(HII or BCD galaxies) is the leakage of photons from HII regions, with shocks 
induced in the ISM by feedback processes also acting as an important ionisation
source.

Finally, our results highlight the importance of accounting for DIG and aperture
effects when interpreting metallicities, particularly in distant galaxies,
where data may be limited to integrated values.
Even in the local Universe, spatially resolved studies based on strong-line
calibrators may be affected by uncorrected DIG contamination, thus potentially biasing
metallicity gradient measurements.

\begin{acknowledgements}
We thank the referee for his/her careful reading of the manuscript
and helpful comments which substantially improved the paper.
PL gratefully acknowledges support by the GEMINI ANID project No. 32240002.
PL (10.54499/DL57/2016/CP1364/CT0010) and TS (10.54499/DL57/2016/CP1364/CT0009) are
supported by national funds through FCT and CAUP. RD gratefully acknowledges support
by the ANID BASAL project FB210003.
\end{acknowledgements}

\bibliographystyle{aa} 
\bibliography{references.bib} 

\end{document}